\documentclass[pra,twocolumn,superscriptaddress]{revtex4}%
\usepackage{amssymb}
\usepackage{amsmath}
\usepackage{graphicx}
\usepackage{dcolumn}
\usepackage{color}
\usepackage{bm}
\usepackage{subfigure}
\usepackage{amsfonts}
\usepackage{appendix}
\providecommand{\U}[1]{\protect\rule{.1in}{.1in}}
\begin{document}
\preprint{APS/123-QED}
\title{Quantum compiling  with diffusive sets of gates}
\author{Y.\ Zhiyenbayev}
\affiliation{Department of Physics, School of Science and Technology, Nazarbayev
University, 53, Kabanbay batyr Av., Astana, 010000, Republic of Kazakhstan}
\author {V.\ M.\ Akulin}  
\affiliation{Laboratoire Aim\'{e}-Cotton CNRS UMR 9188,
L'Universit\'{e} Paris-Sud et L'\'{E}cole Normale Sup\'{e}rieure de Cachan, B\^{a}t. 505, Campus d'Orsay, 91405 Orsay
Cedex, France}
\affiliation{Institute for Information Transmission Problems of the Russian Academy of
Science, Bolshoy Karetny per. 19, Moscow, 127994, Russia}
\author{A.\ Mandilara}
\affiliation{Department of Physics, School of Science and Technology, Nazarbayev
University, 53, Kabanbay batyr Av., Astana, 010000, Republic of Kazakhstan}
\begin{abstract} 
Given a set of quantum gates and a target unitary operation, the most elementary  task of quantum compiling is the identification of a sequence of the gates  that approximates the target unitary to a determined precision $\varepsilon$.
Solovay-Kitaev theorem provides an elegant  solution which is based on the construction of  successively tighter  `nets'    around the unity comprised of successively longer sequences of gates. The procedure  for the construction of the nets, according to this theorem, requires accessibility to  the inverse of  the gates as well. In this work, 
we propose a method  for constructing  nets around  unity without this requirement. The algorithmic procedure is applicable   to sets of gates which are diffusive enough, in the sense that sequences of moderate length  cover the space of unitary matrices in a uniform way.  We prove that the number of gates sufficient for reaching a precision $\varepsilon$ scales  as 
 $  \log  (1/\varepsilon )^{\log 3 / log 2} $ while the pre-compilation time is increased  as compared to thatof the Solovay-Kitaev algorithm by the exponential factor $3/2$.

\end{abstract}
\maketitle

Approximation up to a given accuracy of an arbitrary unitary transformation by a series of standard
transformations is an important ingredient of programming of quantum computers, which was formulated and solved 
\cite{Solovay, Kitaev} in the case where the set of $\mathcal{M}$ predetermined standard transformations contains both direct operations and their inverses. The so called Solovay-Kitaev (SK) theorem 
provides the proof of existence together with the method for constructing the solution. Based on the elements in the proof of the SK theorem, the Dawson-Nielsen
(DNSK) algorithm \cite{Nielsen} provides the exact steps for identifying a
series of length $L$, which scales with the required accuracy $\varepsilon $ as $O \left (\log  (1/\varepsilon )^{3.97}\right )$, and with running time as $O \left (\log  (1/\varepsilon )^{2.71}\right )$. For the special case of qubits, different techniques have been suggested \cite{Horsman, Svore} improving the running
time of this algorithm, while in the general case it has been proved \cite{Shen,Mosca} that the use of extra ancilla qubits further improves the relations
of both the length and the running time, with the accuracy $\varepsilon $. 

Here we address the question \cite{Nielsen} whether it is possible to generalize the results of the SK theorem onto the
case where the set of  predetermined operations does not contain the inverses. 
 In view of the fast development of quantum technologies, this problem
has theoretical  but also  practical interest since experimentalists sometimes do not have access to inverse operations.
%-- they are restricted  to semigroup  rather than group operations. 
For instance, time is the main quantum control (positive) parameter and one  may employ it to construct both a gate and its inverse. On the other hand decoherence effect inducing constraints in time might prevent one from doing so in practice.

\begin{figure}[h]
 {\includegraphics[width=0.4\textwidth]{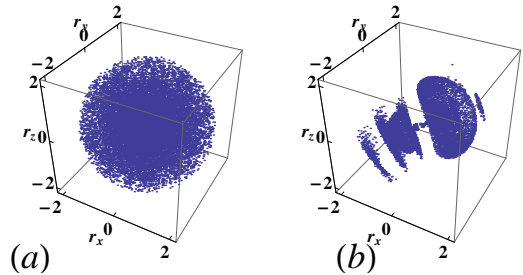}}
\caption{ Geometric representation  of all sequences $\widehat{T}$ of length $17$ generated by two different computationally universal sets of (two) single-qubit gates. \textit{(a)} A diffusive set, and \textsl{(b)} a non-diffusive one. }
 \label{FIG2}
\end{figure}

  Progress on the possibility of extending the SK theorem has   been reported in \cite{Linden, comment} and also in a very recent related work \cite{Ozols}. Our answer  is also positive and conditional on a specific property  of the given set.  We require that sequences of gates of moderate length (composed of $~15-20$ gates) cover the space of unitary matrices in a uniform way. 
This property of sets  was initially investigated in \cite{Arnold} and criteria have been formally developed in \cite{Lubotsky} in the case where the inverse operators  are included in the set. More recently the powerfulness of such sets, so-called `efficiently universal', over just computationally universal ones, has been demonstrated in \cite{Harrow}  for the problem of quantum compiling. In view of lack of formal criteria for the case where the inverses are not available, we  avoid the use of the term of `efficiently universal sets' and we  employ instead the loosely  defined term  of `diffusive sets'. As in \cite{Harrow}, we require that such sets are composed of non-commuting  computationally universal gates and in addition that sequences of \textit{moderate } length composed of the gates of these sets  cover densely and uniformly the space of unitary matrices.
For  the special case of qubits where the property of diffusivity   can be  visually over viewed (see Fig.\ref{FIG2}),  we have found out that a considerable  number of computationally universal sets are also diffusive  (our estimation $\approx 30\%$). In addition, we have noticed that
 by multiplying a set of computationally universal gates with a random unitary  matrix \cite{Zyc}, one   transforms with high probability the former into  a diffusive set.  Physically this random matrix may stand for the free evolution of the quantum system that interpolates the  control actions which generate the  gate operations. %Finally, since there is no visual representation for higher dimensions we can only conjecture that these preliminary numerical observations hold true in higher dimensional Hilbert space. 

Let us now briefly present  the idea of our approach and the structure of this manuscript, assuming that the reader is a little familiar with the proof of the SK theorem.  As in the latter, our aim is to develop a universal  algorithmic method for constructing a series of successive   $\left (\varepsilon _{i} ,\varepsilon _{i}^{2}\right ) -$ nets around the identity and with the requirement $\varepsilon _{i+1}=\varepsilon _{i}^2 $ \cite{comment2}. A $\left (\varepsilon _{i} ,\varepsilon _{i}^{2}\right ) -$ net signifies that in the $\varepsilon _{i} -$ neighborhood of the identity operator there are sequences of gates of length $L_{i}$ and for each of these sequences there is another point at distance less than $\varepsilon _{i}^{2}$. 
 After we introduce our notation and a geometric picture that permits us to interpreter the nets and our methods in a geometric fashion, we explain how the nets  can be  used  to improve the approximation for given $\widehat{U}$ in the standard recursive way. Then we present our main result, a method for successively producing the nets via  `shrinking'. We justify the limits of this method using the theory of random walks and we  confirm our theoretical predictions with  a compiling example for phase rotation gates. 

 \vspace{1cm}

Throughout this work we  consider that   $\mathcal{M}$, the set of given gates, contains just two unitary operations $\widehat{A} =\exp  \left [ -i \widehat{O}_{0}\right ]$ and $\widehat{B} =\exp  \left [ -i \widehat{O}_{1}\right ]$ determined by two Hermitian operators $\widehat{O}_{i =0 ,1}$ in a Hilbert space of $\mathcal{N}$ dimensions. Each sequence of $k$ transformations, i.e. of length $L=k$, picked from the set $\mathcal{M}$ can be encoded by a binary number $\mu  =\left \{f ,j ,\ldots  p ,q\right \}$ in $k$ registers
\begin{equation}\widehat{T}_{f ,j ,\ldots  p ,q} =e^{ -i \widehat{O}_{f}} e^{ -i \widehat{O}_{j}} \ldots  e^{ -i \widehat{O}_{q}} e^{ -i \widehat{O}_{p}} \label{EQ.Series}
\end{equation}where $f ,j ,\ldots  p ,q =0 ,1$. One may attribute a $d=\mathcal{N}^{2} -1$ dimensional real vector $\overrightarrow{r} =\left \{r_{n}\right \}$ to such a sequence $\widehat{T}$, and in general to any unitary transformation $\widehat{U}$, and this geometric representation is particularly useful for our analysis.  The `mapping' can be achieved by casting $\log  \widehat{U}$ in the sum of the $s u (\mathcal{N})$ traceless generators $\widehat{g}_{n}$:
\begin{equation}-\mathrm{i}\log  \widehat{U} = \sum _{n =1}^{d}r_{n} \widehat{g}_{n} = \overrightarrow{r} \cdot \overrightarrow{g}.\label{ve}
\end{equation}
If the generators are normalized,  $\mathrm{Tr}\left(\widehat{g}_{i}.\widehat{g}_{j}\right)=\delta_{ij}$, then  $r_{n}=\mathrm{Tr}\left(-\mathrm{i}\log  \widehat{U}.\widehat{g}_{n}\right)$. The unity operator corresponds to $\overrightarrow{r}=0$, while all other unitary operations corresponds to points in the hyper-sphere around it. Some additional information about this mapping is provided in the Appendix and hereafter we sometimes refer to a unitary $\widehat{U}$  as to a \textsl{point},  implying  the edge of the corresponding vector $\overrightarrow{r}_{\widehat{U}}$  in the $d$-dimensional space. For single qubit operations the a vector $\overrightarrow{r}=\left(r_x,r_y,r_z\right)$ is $3$-dimensional and  can be visualized, and in Fig.~\ref{FIG2} we use this possibility to observe the difference  between a diffusive and a non diffusive set of computationally universal pairs of gates. 

The representation of unitary operators as vectors leads  naturally to the following definition of  \textsl{distance} $\mathcal{D}$ between  two unitary operators: 
\begin{equation} \mathcal{D}\left(\widehat{U}_{1},\widehat{U}_{2}\right)=\left|\overrightarrow{r}_{\left(\widehat{U}_{1}^{-1}.\widehat{U}_{2}\right)}\right| \label{d}\end{equation} where the right-hand side of the equation describes the length of the vector for the unitary matrix $\widehat{U}_{1}^{-1}.\widehat{U}_{2}$. By definition the following property holds $\mathcal{D}\left(\widehat{U}_{1},\widehat{U}_{2}\right)=\mathcal{D}\left(\widehat{U}_{1}.\widehat{U}_{2}^{-1},\widehat{I}\right)$.  In our proofs  where we are mostly  interested in the regime of \textit{small} distances, we use $\mathcal{D}$  as a   measure of distance between unitary operators --in fact we have tested that this perfectly correlates with other  measures in use \cite{Kitaev, Fowler}.

 \vspace{1cm}

Using the introduced notation we can more clearly state now  the objective of the quantum compiling task and the utility of constructing successive nets.\\
\textsl{ Quantum Compiling:} Given an arbitrary unitary transformation $\widehat{U}$, identify a sequence $\widehat{T}_{f ,j ,\ldots  p ,q}$ of gates from the set $\mathcal{M}$, of a total length $L$,  which approximates $\widehat{U}$ to a given accuracy $\varepsilon $, or else,
\begin{equation} \mathcal{D}(\widehat{T}_{\substack{\underbrace{f ,j ,\ldots  p ,q} \\ L}}, \widehat{U}) <\varepsilon . \label{Prod} \end{equation}

Different strategies can be in principle designed to solve this problem and each of them is characterized by three  relations:
the relation between the total length $L$ of the sequence and the achieved precision $\varepsilon $, the relation between the running time of the algorithm and $\varepsilon $, and the pre-compilation time. All these relations cannot simultaneously scale  in an optimal
way  and there is an apparent  interplay among them. For instance, the simplest strategy is the exploration of all possible sequences of a given
length and identification of a sequence closest to the required transformation $\widehat{U}$. \ This is the well-known coverage problem, typical of coding theory, \ which
yields indeed the shortest sequence $L \propto \log  (1/\varepsilon )$ for a given accuracy. However, the identification of such a sequence requires a  time consuming work of exploration of all possibilities, whereas the running time scales exponentially with
$L$ thus making the approach intractable in the high accuracy limit (however see \cite{Fowler} for an enhanced protocol). This strategy does not require pre-compilation time since every
new $\widehat{U}$ requires a new search but on the other hand, and for the same reason, this is not a universal strategy. 

The SK theorem for sets including inverses offers \cite{Nielsen} a balance between the three relations, which could be possibly optimal; both the length of sequence and running time scale poly-logarithmically with $\varepsilon $ and notably these are independent of the dimension of the Hilbert space $\mathcal{N}$, while the pre-compilation time scales polynomially with $\varepsilon $ and exponentially with $d=\mathcal{N}^{2} -1$. The SK theorem, and in consequence DNSK algorithm, are heavily based  on a `successful' construction of a sequence of $\left (\varepsilon _{i} ,\varepsilon _{i}^{2}\right ) -$nets around the identity. Once these nets are constructed and stored, one can perform a standard procedure (ignoring always the telescoping step and assuming $\varepsilon _{i+1}=\varepsilon _{i}^2 $) to approximate a given unitary $\widehat{U}$:
\renewcommand{\labelitemi}{\textperiodcentered}
\begin{itemize}

 \item One  first performs a number of relatively short
sequences  of transformations $\widehat{T}_{\substack{\underbrace{f ,j ,\ldots  p ,q} \\ r}}$  of length $\approx 16-20$ that  serve as initial reference points. Let us call this net \textsl{ the sampling net}.
\item For the given $\widehat{U}$ one has to identify by exhaustive search in the sampling net the closest reference point $\widehat{T}^{\left (0\right )}$ such that  distance $\mathcal{D} (\widehat{U}^{-1}\widehat{T}^{\left (0\right )} ,\widehat{I})<\varepsilon _{0}$. If it is not the case one should restart augmenting the length $r$. 
\item One can then use the net $\left (\varepsilon _{0} ,\varepsilon _{0}^{2}\right )$ in order to identify the sequence $\widehat{T}^{\left (\varepsilon _{0}\right )}$ such as \ $\mathcal{D} (\widehat{U}^{ -1} \widehat{T}^{\left (0\right )} \widehat{T}^{\left (\varepsilon _{0}\right )} ,\widehat{I}) <\varepsilon _{0}^{2}$. 
\item The procedure is repeated $n$ times and at the last step one arrives at  the desired result: a sequence of gates $\widehat{A}$ and $\widehat{B}$ \begin{equation*}\widehat{T}^{\left (0\right )} \widehat{T}^{\left (\varepsilon _{0}\right )} \ldots  \widehat{T}^{\left (\varepsilon _{n}\right )}
\end{equation*} of length $r +\sum _{i=0}^{n}L_{i}$ that reproduces the given unitary in $\varepsilon=\varepsilon_{n}^{2} $ approximation: $\mathcal{D} (\widehat{U}^{ -1} \widehat{T}^{\left (0\right )} \widehat{T}^{\left (\varepsilon _{0}\right )} \ldots  \widehat{T}^{\left (\varepsilon _{n}\right )} ,\widehat{I}) <\varepsilon $. 
\end{itemize}
The dependence of the final length $L$ with $\varepsilon $ is determined by the relation between $L_{i +1}$ and $L_{i}$. If $L_{i +1} =M L_{i }$ where $M \in \mathbb{N}$, \ it is straightforward to prove the dependence is the
desired, poly-logarithmic one:
\begin{equation}L_{n} =r \left (\log  \left (1/\varepsilon _{n}\right )/\log  \left (1/\varepsilon _{0}\right )\right )^{\log  M/\log  2} . \label{Len}
\end{equation}

 \vspace{1cm}

Now, let us turn to the main question of how to construct the sequence of nets and  let us  assume hereafter that the given set  of gates (including the inverses or not) is a diffusive one. The latter condition permits us  to consider  the points on  these nets --even on the sampling one, as uniformly  distributed  according to the Haar measure. Using then the formula for the volume of a sphere in the $d$ dimensional space, one can calculate that the number of required points for a sufficient density is up to a constant factor $\left (\varepsilon _{i}\right )^{ -d}$.
We desire to employ the $K_{i} \propto\left (\varepsilon _{i}\right )^{ -d}$ points/sequences of the $\varepsilon _{i}$-net to identify the $K_{i+1} \propto\left (\varepsilon _{i}\right )^{ -2d}$ points of the consecutive $\varepsilon _{i+1}$-net.

\begin{figure}[h]
 {\includegraphics[width=0.4\textwidth]{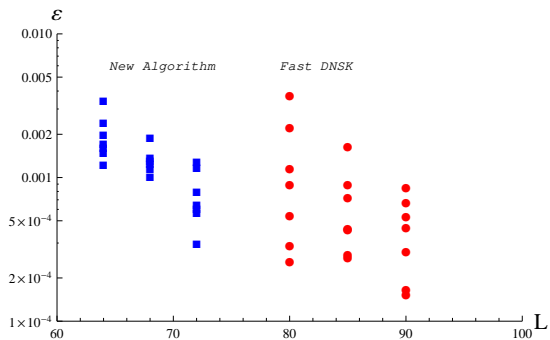}}
\caption{Accuracy of approximation $\varepsilon$ of phase rotation gates $R_{2^d}$ for $d=1,\cdots,7$ by sequences of two diffusive gates, plotted versus  length of the sequence. Blue squares:  results obtained with the introduced algorithm. Red circular dots: results of the Fast DNSK. Different `columns' correspond to different initial lengths $r$ of the sequences  in the sampling net.  From left to right: $r=16, 17, 18, 16, 17, 18$.  }
 \label{FIG1}
\end{figure}

We first consider the case where the inverses are available in the set,  we follow the main idea introduced in \cite{Solovay, Kitaev}, in order to arrive in a simplified version of the DNSK algorithm. The key idea in \cite{Solovay, Kitaev} is to apply
the \textit{ normal commutator} on a pair of  sequences, $\widehat{T}_{\left (1\right )}^{\left (\varepsilon _{i}\right )}$ and $\widehat{T}_{\left (2\right )}^{\left (\varepsilon _{i}\right )}$ of the $\varepsilon _{i}$-net, employing also the inverses of these, $\left.\widehat{T}_{\left (1\right )}^{\left (\varepsilon _{i }\right )}\right.^{ -1}$ and $\left.\widehat{T}_{\left (2\right )}^{\left (\varepsilon _{i }\right )}\right.^{ -1}$, which are naturally included in the same net. By definition a normal  commutator is $\widehat{T}_{\left (1\right )}^{\left (\varepsilon _{i }\right )} \widehat{T}_{\left (2\right )}^{\left (\varepsilon _{i }\right )} \left.\widehat{T}_{\left (1\right )}^{\left (\varepsilon _{i }\right )}\right.^{ -1} \left.\widehat{T}_{\left (2\right )}^{\left (\varepsilon _{i }\right )}\right.^{ -1}$ and the result of this product is a new sequence at distance less than $\varepsilon _{i}^{2}$ from the unity. This new point/sequence  can be included  in the consequent $\varepsilon _{i+1}$-net. The normal commutator thus naturally leads to the so called `shrinking' of the initial net. The number of distinct normal commutators that can be formed by $K_{i}$ points of the $\varepsilon _{i}$-net matches the number of points required in the  $\varepsilon _{i+1}$-net, independently of the dimension of the Hilbert space and there is no need for additional  `search' steps during the pre-compilation stage. Now concerning the scaling of the length of the sequences with the $i$-th order of the net, $L_{\varepsilon _{i +1}} =4 L_{\varepsilon_i }$ and by inserting $M =4$ into Eq.(\ref{Len}) we arrive to a quadratic dependence between length and accuracy: $L_{n} \sim \left (\log  \left (1/\varepsilon _{n}\right )\right )^{2}$. In what has been described we  have   ignored     the extra step of  `telescoping' \cite{comment2} and we name this simple and faster version  \textit{fast DNSK}. This simplified  version can be compared with the algorithm that we suggest on the same grounds. In addition the requirement of diffusive characters of gates seems to partially compensate for the `telescoping' procedure (see standard deviation of the approximation in Fig.~\ref{FIG1}).

Now, let us consider the case where the inverses are not accessible and therefore the idea of normal commutators is not-applicable.  
%We suggest here an alternative shrinking procedure based on a diffusion process using the points of the $\varepsilon _{i}$-net, followed up by post-selection.
Let us  start as before with the $\varepsilon _{i}$-net and  select at random $M$ sequences from this net. 
Then construct a new sequence by taking one  of the ($M!$) products of these sequences. 
If $\varepsilon _{i}$ is small enough  one may interpreter, in approximation $O(\varepsilon _{i}^2)$, this new point as a result of an $M$-steps random walk in the $d$ dimensional space (see Lemma in the Appendix). More precisely the steps of this random walk are the vectors/sequences of the $\varepsilon _{i}$-net, which are isotropic in the  $d$ dimensional space and their size is in the interval $[0,\varepsilon]$ with a standard deviation of the order $~\varepsilon$ as well.
If now all $M$ products are produced from the sequences/points of the $\varepsilon _{i}$-net, the resulting $\left (\varepsilon _{i}\right )^{ -M d}$ new points are going to follow the distribution of a random walk and diffuse out of the unity.
Such  random walks have been well studied (see, for instance, \cite{MIT}) and it is straightforward to derive  the probability of finding a new point/sequence at distance $r=\left|\vec{r}\right|$ from the origin of the hyper-sphere after $M$ steps,
\begin{equation}
P_M(r)=2 \left(\frac{d}{2 M  \varepsilon^2}\right)^{d/2}\frac{r^{d-1}}{\Gamma(\frac{d}{2})}\mathrm{e}^{-\frac{d r^2}{2 M  \varepsilon^2}}. \label{dist}
\end{equation}
To build the $\varepsilon _{i+1}$-net one needs  to post-select from the new diffused distribution of points/sequences  the ones at distance  $\mathcal{D}<\varepsilon _{i}^2$ from the origin.

To claim that the suggested  method for shrinking is applicable  we need though to answer  three questions: \\
\textit{(a)} What is the minimum number of steps $M$ that provides the required density of points for the consequent $\varepsilon _{i+1}$-net?\textsl{(b)} How does the time for constructing the nets  compare to the pre-compilation time of fast DNSK? \textsl{(c)} Is the quality of the produced $\varepsilon _{i+1}$-net good enough to ensure the successful construction of the $\varepsilon _{i+2}-$net?

It turns out   that $M=3$ gives sufficient density of points inside the radius $\varepsilon _{i}^2$ for any dimensions $d$. To prove this statement we first calculate the cumulative distribution function for probability distribution Eq.(\ref{dist}), plug in $\varepsilon _{i}^2$ and arrive to
\begin{equation} 
P_M(r<\varepsilon _{i}^2)=1-\frac{\Gamma\left(\frac{d}{2},\frac{d \varepsilon _{i}^2}{2 M}\right)}{\Gamma\left(\frac{d}{2}\right)}. \label{cudist}
\end {equation}
However an approximate formula for the distribution in Eq.~(\ref{cudist}) is more easy to overview as
\begin{equation} 
P_M(r<\varepsilon _{i}^2)\approx 2 \left(\frac{d}{2 M  }\right)^{d/2}\frac{\varepsilon^{d}}{\Gamma(\frac{d}{2})}. \label{cuappro}
\end {equation}
The latter can be derived by noting that the maximum of the `Rayleigh type' of distribution in Eq.~(\ref{dist}) is outside the region of $r<\varepsilon _{i}^2$ and thus its contribution can be ignored.
If Eq.(\ref{cuappro}) is  multiplied with the total number of points $\left (\varepsilon _{i}\right )^{ -M d}$ resulting from the diffusion process, one arrives at a formula that provides the number of points at distance less than $\varepsilon _{i}^2$ from the origin (unity). For  $M=3$,
the required (for sufficient density) order   $\left (\varepsilon _{i}\right )^{ -2 d}$ is reached for any dimension $d$.

Since $M=3$ is sufficient one needs to perform all triplet products of the sequences in the $\varepsilon _{i}$-net and then post-select the points/sequences for which  $\left|\overrightarrow{r}\right|<\varepsilon_i^2$. The number of sequences of  the $\varepsilon _{i}$-net is approximately $K_{i} \propto\left (\varepsilon _{i}\right )^{ -d}$ and therefore the number of necessary operations is $\propto\left (\varepsilon _{i}\right )^{ -3d}$. In the fast DNSK the number of operations for constructing all normal commutators is $\propto\left (\varepsilon _{i}\right )^{ -2d}$. We may thus conclude that the time in our method is increased exponentially by a  factor of $1.5$. We believe that this is a natural consequence of the fact that the two methods have the same running time, but our suggested method achieves better scaling of length with approximation (see Eq.(\ref{Len}) with $M=3$ while for fast DNSK $M=4$). In the Appendix we additionally prove that it is very unlikely that the long pre-compilation time of the algorithm that we suggest here can be shortened. There we show that if the post-selection process on the points is replaced by a pre-selection process, then this problem maps into an $NP$ hard problem, namely, to the $0-1$ Knapsack problem in $d$ dimensions. 

Addressing now the last question. The number of points inside the radius $\varepsilon _{i}^2$ is increasing as $r^{d-1}$ (see Eq.(\ref{dist})) and therefore has approximately  the desired dependence of a uniform distribution. In addition, under our assumption of diffusive set,  the new points should be distributed in an isotropic way. Here though we suggest an additional step to ensure isotropicity which we have found very useful in practice:   for each point/sequence identified to belong in the $\varepsilon _{i+1}$-net construct \textit{all  cyclic permutations} of the gates in the sequence. Cyclic permutations leave the spectrum of the sequence intact, and  the length of the vector $\left|\overrightarrow{r}_{\widehat{U}}\right|$  depends only on the spectrum of the corresponding unitary $\widehat{U}$. Therefore, cyclic permutations of sequence leave the distance $\mathcal{D}$ from unity invariant and the corresponding new points are distributed over the hyper-spherical surface of the original point. 
 \vspace{1cm}

In the Fig.~\ref{FIG1} we present   quantum compiling results obtained with the proposed algorithm versus the fast DNSK. More precisely, we approximate the phase rotation gates,
\begin{equation}
R_{2^d}= \left( \begin{array}{cc}
1& 0 \\
0 & \mathrm{e}^{\mathrm{i} \pi/2^d} \end{array} \right),~~ \mathrm{with}~ d=1,\ldots 7
\label{rot}\end{equation}
using the introduced algorithm and then the fast DNSK keeping the parameters of produced nets very similar in both cases. For both methods we have used the same pair of diffusive gates (see Appendix), but naturally for the latter we have included the inverses. On each  `column' the seven points describe the approximation of the seven phase rotations Eq.(\ref{rot}). There is no correlation between the precision achieved and the order $d$ of the phase gates and for this reason we 
have not marked with $d$ the points on the plot. For each method we present three numerical results (three columns) that correspond to three different lengths  of the initial sampling net $r=16, 17, 18$, giving different lengths $L$ to the final sequence that approximate the gate (horizontal axis on  Fig.~\ref{FIG1}).  
For both methods we have used the sampling  and the $\varepsilon_0$  nets.
 To quantify the accuracy $\varepsilon$ we use as measure of distance: $ d_F \left(\widehat{U}_{1},\widehat{U}_{2}\right)=\sqrt{\frac{2-\left| \mathrm{Tr}\left( \widehat{U}_{1}\widehat{U}_{2}^{-1}\right)\right|}{2}}$, introduced in \cite{Fowler}. More technical details on this example can be found in the Appendix while the related programs can be downloaded from the site www.qubit.kz.

The numerical results in Fig.~\ref{FIG1} confirm our theoretical prediction that the suggested algorithm provides better scaling of length with accuracy than the DNSK.
On the contrary from the graph one cannot extract the  scaling of the length with accuracy, Eq.(\ref{Len}). This would require results where different orders of nets
are used;  here we only change the parameters of the sampling net and we use  the first net around unity for all the results. The accuracy achieved for different gates is not uniform because we do not employ a procedure for extracting extra points which would  improve the quality of nets in terms
of homogeneity. Therefore we think that our suggested algorithm can be further upgraded by adding this additional procedure and possibly other procedures which would extend its applicability to sets of gates which are not completely  but close to being diffusive.

 \vspace{1cm}

In conclusion, we have suggested an algorithmic procedure for generating  nets of sequences of gates around  unity  under the condition that the given sets of gates are diffusive. This algorithm results in better scaling of length with the approximation than a simplified fast version of the DNSK algorithm does, and works in both  the presence and the absence of inverses. The improvement in scaling can be  justified by the fact that there is an exponential counter-increase in pre-compilation time, as  compared to the DNSK algorithm. This confirms an expected interplay between the relations characterizing  algorithmic procedures  solving the same problem. When the inverses are included in the set, the  notion of diffusive sets    converges to the notion of `efficiently computational sets' introduced in \cite{Harrow} and our results  partially fulfill the predictions of that work concerning the considerable improvement of the scaling of length with accuracy. Finally, the accurate characterization of the diffusive property of a set of gates remains an interesting open question, deserving further investigation. 

%The suggested algorithm is applicable in practice only to compilation of  $SU(2)$ operations by diffusive sets of single qubit gates. It remains an open question whether the algorithm can be further modified to be practical for higher dimensional Hilbert spaces. However if no further assumptions are taken, this seems rather unlikely to us as we prove in this work that this problem maps to  $0-1$ Knapsack problem in $d$ dimensions. Therefore our results rather support the idea that an extension of Solovay-Kitaev theorem to semigroups is possible in theory but rather unachievable in practice, a conclusion that has been drawn independently in \cite{Linden}.

%For higher dimensions, one may solve this problem with the help of $\mathcal{N}$ free parameters in  auxiliary  local qubit gates. More precisely, if one adds to the set of   gates $\widehat{A}$ and $\widehat{B}$ $\in SU(\mathcal{N})$,   single qubit gates with $\mathcal{N}$ free control parameter, then the unity can be `built' in a efficient way \cite{Harel}. Departing from the unity, one may obtain pairs of { gate, inverse of gate} and then proceed with the SK algorithm.   This idea is to be further elaborated  in a future work.

\section*{Acknowledgment}
AM is thankful to M. Lukac for bringing  this problem to her attention. AM and VA are thankful to G. Kabatyanski for pointing out the connection to the knapsack problem  and to M. S. Byrd for many  useful  initial discussions. The authors are grateful to A. Harrow for his  comments  which have  helped them to draw conclusions from preliminary results and to the unknown referee
for much valuable feedback. AM and YZ  acknowledge financial support during this work from the Ministry of Education and Science (MES) of the Republic of Kazakhstan via the contract $\#$ $339/76-2015$. AM also acknowledges financial support from the Nazarbayev University ORAU grant ''Dissecting the Collective Dynamics of Arrays of Superconducting Circuits and Quantum Metamaterials''  and MES RK state-targeted program BR$05236454$.

\appendix
\section{Additional information on the geometric representation of unitary matrices}

 The representation of a unitary $\widehat{U}$ via $\overrightarrow{r}$ in general ignores the global phase since $\widehat{g}_{0}=\widehat{\mathrm{I}}$ is excluded from the set of generators. On the other hand one obtains  different vectors for $\widehat{U}$ and $-\widehat{U}$ and in this work  we want to totally ignore the global phase. This problem can be  resolved  when the $\mathcal{N}$ dimensional quantum system stands for  an assembly of $n$ qubits,   ($\mathcal{N}=2^n$). In this case, the introduced geometric space of unitaries  is a  $d$-dimensional hyper-sphere of radius $\left|\overrightarrow{r}_{\max}\right|=2^{\frac{n}{2}} \pi$ centered at $\left|\overrightarrow{r}_{\min}\right|=0$. The unitary $-\widehat{I}$ is located on the outer surface while $\widehat{I}$ in the center of hyper-sphere. This unwanted  discrepancy can be corrected by the following mapping:     if $\left|\overrightarrow{r}\right|>2^{\frac{n}{2}-1}\pi$ then $\overrightarrow{r}\longrightarrow -\overrightarrow{r} \left(2^{\frac{n}{2}}\pi-\left|\overrightarrow{r}\right|\right)/\left|\overrightarrow{r}\right|$. 

We add here a \textsl{Lemma} that  and which can be  proved using the Baker Campbell Hausdorff formula:
\renewcommand{\labelitemi}{\textendash}
\begin{itemize}
	\item if $\left|\overrightarrow{r}_{\widehat{U}_1}\right|<\varepsilon$ and $\left|\overrightarrow{r}_{\widehat{U}_2}\right|<\varepsilon$ where $\varepsilon <<1$  then 
 \begin{equation} \overrightarrow{r}_{(\widehat{U}_1.\widehat{U}_2)}=\overrightarrow{r}_{\widehat{U}_1}+\overrightarrow{r}_{\widehat{U}_2}+O(\varepsilon^2).\label{lemma}\end{equation}
\end{itemize}

\section{Details on the example}
To generate  Fig.~\ref{FIG1} we  used a diffusive set $\mathcal{M}$ composed of the gates $\left\{\widehat{A},\widehat{B} \right\}$. More precisely, 
$\widehat{A}=\widehat{ H}\cdot \widehat{F}$ and $\widehat{B}= \widehat{T}\cdot \widehat{F}$ where $\widehat{H}$ is the Hadamard gate, $\widehat{T}$ the $T$-gate and $\widehat{F}$ a randomly generated unitary matrix
\begin{equation}
\widehat{F}= \left( \begin{array}{cc}
-0.40194 - \mathrm{i} 0.43507 & -0.36803 - \mathrm{i} 0.71674 \\
0.36803- \mathrm{i} 0.71674 & -0.40194 + \mathrm{i} 0.43507  \end{array} \right).
\end {equation}
For the fast DNSK we have used the analogous set $\mathcal{M'}$ composed of the gates $\left\{\widehat{A},\widehat{B},\widehat{A}^{-1},\widehat{B}^{-1} \right\}$.

To achieve the approximations to the phase rotation gates $R_{2^d}$ for the case of the algorithm based on diffusion, we perform the following  steps:
\begin{itemize}
	\item We create all the sequences of length $L=16$.  This is the sampling net composed by $k=2^{16}$ points.
	\item  From this  net covering all space we select the points inside the radius $\varepsilon _{s}=0.3$. The  $\varepsilon _{s}$ is calculated according to the formula
	\begin{equation}
	\varepsilon _{s}=\frac{2^{1/4}\sqrt{\pi}}{k^{1/3}}.
	\end{equation}
	\item We perform the `diffusion' process creating all triplets, and then we post-select the ones which are inside the radius $\varepsilon _{0}=(\varepsilon _{s})^2$. We add $45$ permutations for each successful sequence. This way we create more points than the ones needed for $\varepsilon _{0}$ so we randomly select from these the sufficient number, $8 /\varepsilon _{0}^3$.
	\item We use the  sampling and the $\varepsilon _{0}$-net  to identify the sequences of total length $65$ that approximate each of the seven gates.
	\item We repeat the procedure for initial lengths $L=17$ and $L=18$, to obtain better approximation with sequences of lengths $68$ and $72$ respectively. 
\end{itemize}
We note here that the whole procedure is very fast since of course the aimed precision is low.

For the fast DNSK algorithm the steps are identical apart from the fact that we construct the normal commutators instead of triple products. For consistency, we include the permutation step.

\section{Pre-selecting instead of post-selecting}

In the main text we have studied the straightforward method for achieving shrinking by performing a diffusion process followed up by post-selection.
More precisely, our suggestion is to  construct all possible triplets from the  $K_{i} \propto\left (\varepsilon _{i}\right )^{ -d}$ points/sequences of the $\varepsilon _{i}$-net, calculate for each of the resulting sequences $\left|\overrightarrow{r}\right|$ and then post-select those  with $\left|\overrightarrow{r}\right|<\varepsilon_i^2$.  Is there a more efficient  way for doing this? Let us first replace  the post-selection  by pre-selection noting the following: \vspace{0.5cm}
\\ \textit{Given  the  sequences $\widehat{T}^{(\varepsilon_i)}_1$,  $\widehat{T}^{(\varepsilon_i)}_2$,  $\widehat{T}^{(\varepsilon_i)}_3$ with corresponding vectors $\overrightarrow{r}_1$, $\overrightarrow{r}_2$ and $\overrightarrow{r}_3$ which satisfy the condition $\left|\sum_{j=1}^3\overrightarrow{r}_j\right|< \varepsilon_i^2$ then  $\mathcal{D}( \widehat{T}^{\varepsilon_i}_1.\widehat{T}^{\varepsilon_i}_2.\widehat{T}^{\varepsilon_i}_3,\widehat{I} )<O(\varepsilon_i^2)$.}\vspace{0.5cm}
\\This pre-selection process on the points of the initial $\varepsilon_i$-net closely resembles  a known computational problem: the $0-1$ Knapsack problem in $d$ dimensions. Let us  briefly state this problem:
\vspace{0.5cm}
\\ \textit{Given $n$ $d$-dimensional vectors $\vec{v}_i$ with positive entries and $p_i>0$ profit for each of them, and a $d$-dimensional bin $\vec{B}$ find the $n$-dimensional vector $\vec{x}$ with $0-1$ entries such that}:
\begin{itemize}
\item  $\sum_{i=1}^{n}x_i p_i$ is maximized 
\item it is subject to $\sum_{i=1}^{n}x_i \vec{v}_i\leq\vec{B}$
\end{itemize}
 The  mapping of the pre-selection problem to the Knapsack problem is almost straightforward  one needs to \textit{(a)} make the entries of input points/vectors $\overrightarrow{r}_i$  from the $\varepsilon_i$-net strictly positive (so that they can represent $\vec{v}_i$), \textit{(b)} attribute a  profit $p_i$ to these vectors  and \textit{(c)}   adjust the entries of the bins $\vec{B}$ to the requirements of the $\varepsilon _{i+1}-$ net.  The first task can be done by adding a fixed vector $\vec{v}_0 $ with $\left|\vec{v}_0 \right|>\varepsilon _{i}$ to all $\overrightarrow{r}_i$. Concerning the profit one can attribute the same profit to all input vectors but instead of maximizing the total cost, just minimize it. Finally, the entries of the bin should be adjusted to  $B_k=3\vec{v}_0+\varepsilon _{i+1}/\sqrt{d}$ for $k=1,\ldots d$ .
 
It has been proven \cite{Frieze} that there is no fully polynomial time approximation scheme for $d$-dimensional knapsack and that this is an NP-hard problem. In addition there is no efficient polynomial time approximation scheme (EPTAS) \cite{Kulik} even for low dimensions as $d=2$. We may conclude that the pre-selection process for our suggested method is not computationally tractable in $d$ dimensions given also the fact that $n$ ($=K_i$ for our case)  increases  exponentially with $d$.


\begin{thebibliography}{99}
\bibitem{comment} In \cite{Linden}  the set of given operations $\mathcal{M}$ is complimented by the so called Weyl operations, which are mutually orthogonal roots of unity. Since the inverse of a root of unity, is just the same operator to some integer power, it seems that this assumption relaxes the condition of no access to  inverse operations.
\bibitem{comment2} In  SK theorem \cite{Kitaev} the sequence of nets follows a `weaker' relation, $\varepsilon _{i+1} =\varepsilon _{i}^{3/2}$, than the  one targeted in this work. The reason, is that for preserving a good quality of nets one needs  to compliment   the `shrinking' process with a `telescoping' one, at some cost.
\end{thebibliography}
\end{document}